\documentclass[a4paper,11pt]{article}
\usepackage{pos}

\newcommand*{\be}{\begin{equation}}
\newcommand*{\ee}{\end{equation}}

\newcommand*{\bma}{\begin{matrix}}
\newcommand*{\ema}{\end{matrix}}

\newcommand*{\la}{\left\langle}
\newcommand*{\ra}{\right\rangle}

\newcommand*{\bea}{\begin{eqnarray}}
\newcommand*{\eea}{\end{eqnarray}}

\newcommand*{\pref}[1]{(\ref{#1})}

\newcommand*{\tl}{\mathrm{tl}}

\newcommand*{\str}{\mathrm{str}}

\newcommand*{\no}{\noindent}

\title{Composite objects in quantum (super)gravity}

\author*[a]{Axel Maas}
\author[a,b]{Simon Pl\"atzer}
\author[a]{Felix Pressler}

\affiliation[a]{Institute of Physics, NAWI Graz,\\
  Universit\"atsplatz 5, 8010 Graz, Austria}
\affiliation[b]{Particle Physics, Faculty of Physics,\\
University of Vienna, Boltzmanngasse 5,\\
A-1090 Wien, Austria}

\emailAdd{axel.maas@uni-graz.at}
\emailAdd{simon.plaetzer@uni-graz.at}
\emailAdd{f.pressler@edu.uni-graz.at}

\abstract{It has been a long entertained idea that self-bound gravitons, so-called geons, could be a dark matter candidate or form (primordial) black holes. The development of viable candidates for quantum gravity allows now to investigate these ideas. Analytic methods show that the description of geons needs to be based on composite operators made out of the graviton field. We present results from a numerical investigation into  this idea using causal dynamical triangulations, an ab-initio non-perturbative definition of quantum gravity based on general relativity, and accessible in lattice-gauge-theory-like simulations. Our results suggest an interesting dependence on cosmological time and other unexpected features. Finally, we extend the analytic part of the setting to a supergravity scenario. This provides hints which, if confirmed, could explain why supersymmetry may in a realistic universe in principle not be observable at low (collider) energy scales.
}

\FullConference{The European Physical Society Conference on High Energy Physics (EPS-HEP2025)\\
7-11 July 2025\\
Marseille, France\\}

\begin{document}

\maketitle

\section{Introduction}

A full, non-perturbatively valid, description of quantum gravity is arguably still one of the central challenges in physics. That issue is aggravated by the absence of unambiguous experimental or observational input, which could be used to discriminate between different options. Among the possible contenders, theories which non-perturbatively quantize the Einstein-Hilbert Lagrangian stand somewhat out as a minimal solution. In particular using dynamical triangulation (DT) simulations \cite{Ambjorn:2024pyv,Loll:2019rdj,Ambjorn:2012jv,Dai:2021fqb} and functional renormalization group results \cite{Reuter:2019byg} a reasonably consistent scenario \cite{Ambjorn:2024qoe,Loll:2019rdj} of quantum gravity has been developed. These features of minimality and the consistency do not, of course, make this scenario more likely to be the correct theory of quantum gravity. But it provides a frame to tackle many conceptual and practical questions non-perturbatively. Therefore, causal dynamical triangulations (CDT), will be used in the following, with the understanding that many of the conceptual insights will likely carry over to different formulations.

Any theory of quantum gravity has to contend with the fact that classical gravity is a gauge theory of translations \cite{Hehl:1976kj}. Consequently, observables in a path-integral quantization need to be invariant under coordinate transformations and can depend only on invariant distances, e.\ g.\ geodesic distances \cite{Ambjorn:2012jv,Maas:2019eux,Schaden:2015wia}. Suitable candidates are curvature invariants \cite{Misner:1973prb}. One option for a discretized curvature invariant in CDT is the quantum Ricci curvature scalar (QRCS) $Q$, from which in the continuum limit of a smooth manifold the usual curvature scalar $R$ can be extracted \cite{Klitgaard:2017ebu,Loll:2023hen}. This allows already to infer information on global properties, which will be discussed in section \ref{s:global}.

Even more interesting is the question of dynamical degrees of freedom. Gravitons, like gauge bosons, are gauge-dependent, and thus not suitable candidates. The analogue to glueballs in pure gravity are geons, or graviballs \cite{Wheeler:1955zz,Perry:1998hh,Pastor-Gutierrez:2024sbt}. Physically, such geons could take the form of (primordial) black holes or could be dark matter candidates \cite{Maas:2019eux}. Thus, obtaining their properties is highly interesting for phenomenological reasons. The propagators of such geons can be accessed in CDT in a similar way as in lattice gauge theory, as the seminal work \cite{vanderDuin:2024pxb} has demonstrated in two dimensions. Results \cite{Maas:2025rug} in four dimensions will be given in section \ref{s:geon}. In particular, hints for a massive-like behavior is found, with a cosmological-time-dependent mass.

Motivated by these successes, a first step beyond Einstein-Hilbert gravity is made in section \ref{s:sugra}, to the simplest supergravity theory with matter. Building upon the insights on the need to remain manifestly invariant together with the assumption that global space-time properties work similar as in section \ref{s:global} allows to infer why supersymmetry becomes unobservable in the same sense as, e.\ g.\ color is not observable.

Some of the many open questions and ways forward are finally summarized in section \ref{s:summary}.

\section{Global properties}\label{s:global}

CDT can be simulated using Monte Carlo techniques. For that, the code of \cite{Ambjorn:2021yvk} is employed, which has been made accessible to us by the authors. The simulations are done on triangulated foliated manifolds, with spherical boundary conditions on each time slice. Details of the simulations can be found in \cite{Maas:2025rug}. The lattice constant has been estimated to $\approx2.1$ Planck lengths \cite{Ambjorn:2008wc}. This setup has been extensively analyzed \cite{Ambjorn:2012jv,Ambjorn:2008wc,Klitgaard:2017ebu}. Our own simulations agree with these results.

\begin{figure}
 \includegraphics[width=0.5\textwidth]{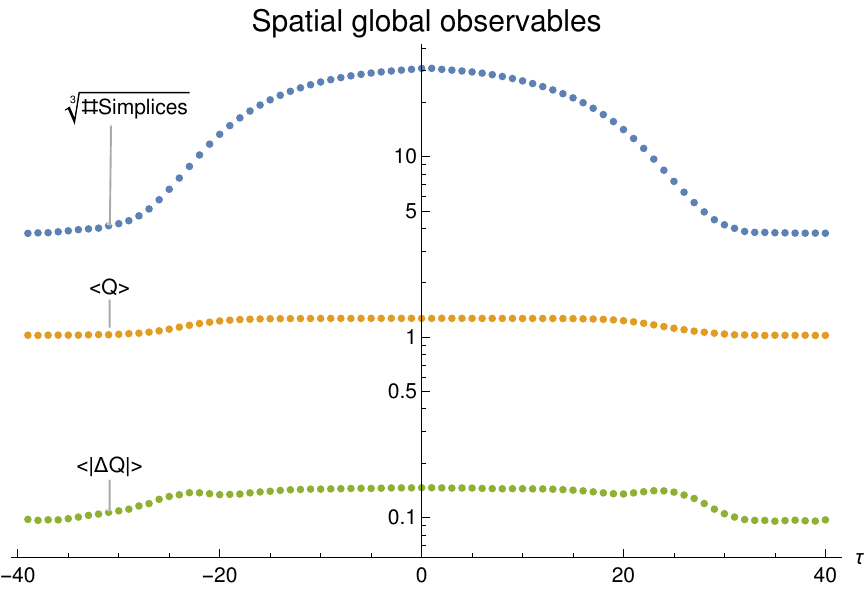} \includegraphics[width=0.5\textwidth]{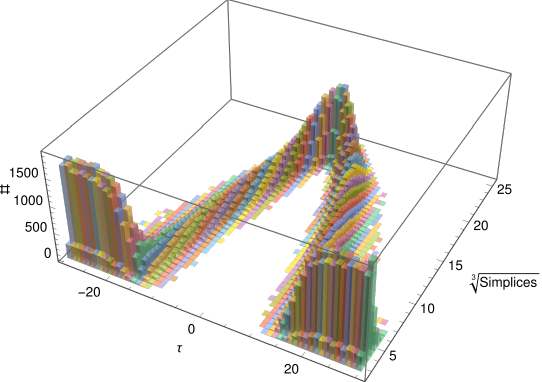}
 \caption{Left: The average global properties \cite{Maas:2025rug} as a function of cosmological time $\tau$: The size in terms of simplices and the value and fluctuation of the quantum Ricci  curvature scalar. Right: The size measured in simplices, but per configuration \cite{MPP:unpublished}. Note that the simulations were done using three-dimensional spherical boundary conditions and periodic temporal boundary conditions. Left panel is from a simulation with 320k simplices and 80 time-slices, right panel from 80k simplices and 60 time slices. Always a smearing value of $\delta=6$ \cite{Maas:2025rug} is used for $Q$.}
 \label{global}
\end{figure}

The most salient features are displayed in figure \ref{global}. Per configuration a cosmological time is introduced, which is measured relative to the largest extent, as measured by the cube root of simplices per time slice. The resulting profile is well-known, and commensurate with an Euclidean de Sitter profile \cite{Loll:2019rdj,Ambjorn:2012jv}. The interesting fact is that the fluctuations around this density profile per configuration are small. This implies that the quantum geometry only fluctuates mildly around the classical de Sitter profile. This is commensurate with the idea that quantum gravity exhibits a Brout-Englert-Higgs effect \cite{Maas:2019eux}, in which in a gauge-fixed language deviations from a vacuum expectation value of the metric would be small.

The fluctuations of the Ricci curvature scalar $Q$ shows consequently only mild fluctuations, which are most strongest during the rapid expansion triggered by the non-zero cosmological constant of a de Sitter-like universe. During this epoch, also the size fluctuations appear largest. Interestingly, the value of this curvature-measure is actually largest in the epoch of largest extent, and not during the rapid expansion or before. This behavior will reflect itself later.

It is interesting, though possible coincidental, that the increase of curvature density towards the maximal extent is also compatible with what could be deduced from the most recent DESI results. There, a time-dependent dark energy is favored \cite{DESI:2024mwx}. This is also seen in EDT calculations \cite{Dai:2024vjc}.

\section{Geon}\label{s:geon}

\begin{figure}
 \includegraphics[width=0.5\textwidth]{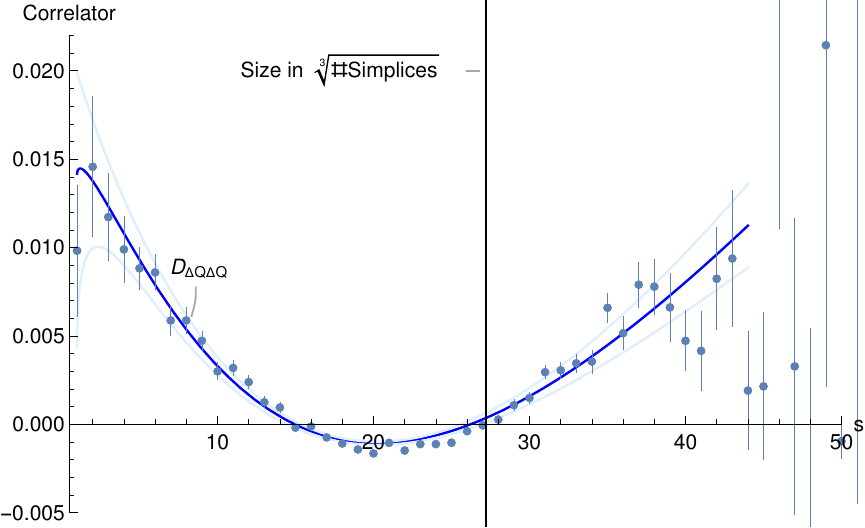} \includegraphics[width=0.5\textwidth]{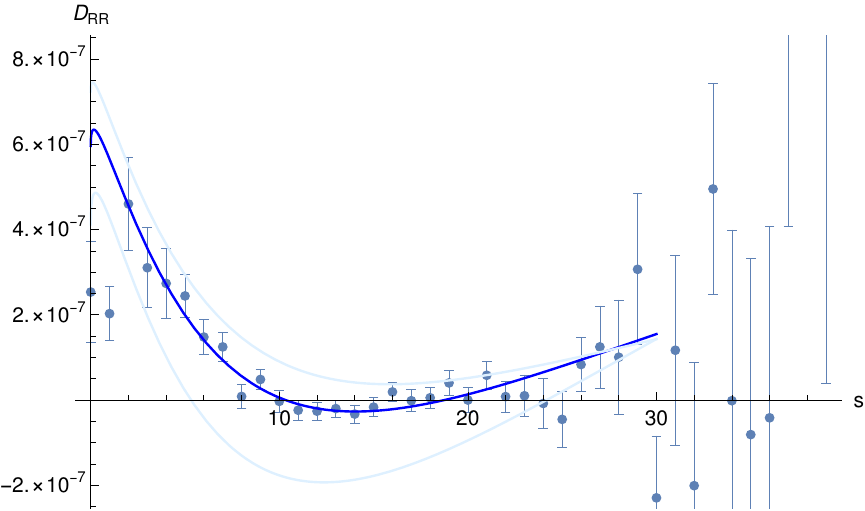}
\begin{minipage}[c]{0.5\textwidth}
 \includegraphics[width=\textwidth]{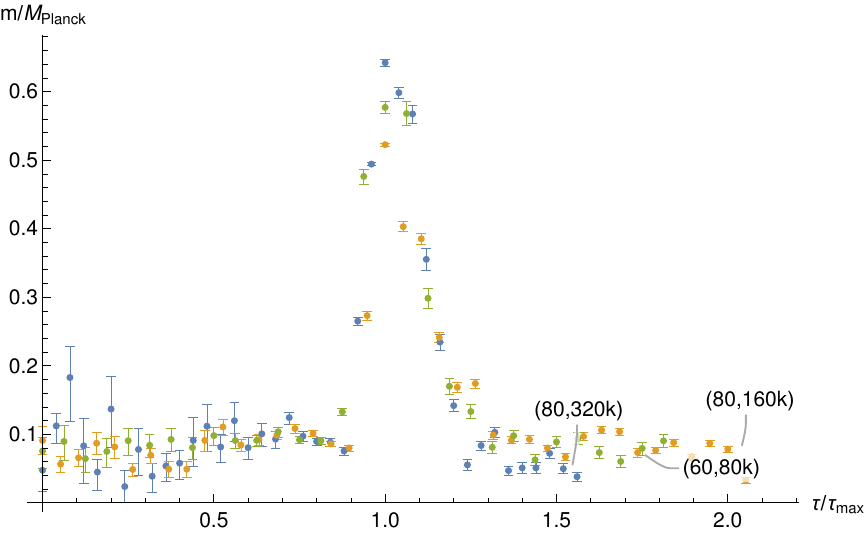}
 \end{minipage}\hfill
 \begin{minipage}[c]{0.4\textwidth}
 \caption{The normalized, connected correlator of the Quantum Ricci curvature scalar (top left), together with a fit of type $a+b\exp(-ms)+cs^d$. Top right panel the same for the extracted curvature scalar, but where the fit is mass $m$ and exponent $d$ fixed to be the same as for $Q$. Lower right panel is the extracted mass for different volumes and time extensions as a function of cosmological time normalized to the peak cosmological time. From \cite{Maas:2025rug,MPP:unpublished}.}
 \label{geon}
 \end{minipage}\hfill
\end{figure}

The results for the connected\footnote{Build from the fluctuation operators $\Delta O(x,\tau)=O(x,\tau)-\int d^3V O(x,\tau)$. I.\ e.\ the spatial fluctuations at fixed cosmological time are considered here to be the fluctuation fields. In flat space-time, this would not make a difference.} correlator of both the QRCS and the extracted ordinary curvature scalar are shown in figure \ref{geon} \cite{Maas:2025rug} as a function of geodesic distance for cosmological time $\tau=0$, i.\ e.\ at maximum extent. Both show a quite similar behavior, though the latter is statistically much more noisy. Their short-distance behavior is compatible with an exponential decay, similar to what would be expected from the behavior in flat space-time. The same is also observed for correlators build from $Q^2$ \cite{Maas:2025rug}. In fact, the effective mass for the exponential part is, within errors, compatible between all three of them \cite{Maas:2025rug,MPP:unpublished}. That all three cases show a similar behavior could be expected if there exist an (intermediate) distance regime in which the behavior is almost the one of flat space-time, and an effective LSZ construction is applicable \cite{Maas:2025rug}. Then, the choice of operator would not matter for the state observed. Together with the exponential behavior expected on space-like separation, this could be a hint for a particle-like excitation, a geon.

It is noteworthy that the lattice is rather coarse with respect to the Planck scale. Thus, short distance artifacts, which may be visible at very short distances, are not expected to be large. At a consequence, the effective volumes are large. There is also an interesting behavior at long distances, where the exponential decay turns into a, within sizable errors, linear rise. This happens at distances which become comparable to the maximal extent of the universes \cite{MPP:unpublished}. Thus, this behavior, which probes a different distance sale, is no longer particle-like. It really describes curvature correlations between distances of the universe of the size of the maximal extent, including possible wrap-around effects due to the spatial spherical boundary conditions.

Taking for the moment the exponential fall-off as indeed originating from a massive-particle-like behavior, the cosmological-time dependence of the mass is also shown in figure \ref{geon}. using the scale of \cite{Ambjorn:2008wc} implies a mass of about 0.1-0.2$M_\text{Planck}$ for the geon, except during the rapid expansion time. There, the mass rapidly increases, simultaneously with the fluctuations of the QRCS. While there is no non-trivial inflation happening this shows that effective degrees of freedom of gravity could very well depend on the expansion rate. If the geon should indeed be related to physical objects like (primordial) black holes, this may hint towards possible new dynamics in the early universe. It is also interesting that after rescaling of the time scale these results are fairly similar between the different simulation sizes. This again emphasizes that volume effects are likely small.

However, many other lattice effects, especially discretization artifacts, may be large. Thus, these results should be taken with caution until more systematics are available. However, they evidently point the way to a large number of new possibilities waiting to be uncovered.

\section{Supergravity}\label{s:sugra}

The results show that a manifestly diffeomorphism-invariant treatment of quantum gravity indeed has very interesting properties. Moreover, together with the hints for the presence of a Brout-Englert-Higgs effect this points towards also a new way of understanding the observational absence of supersymmetry \cite{Maas:2023emb}, despite its key role in unifying gravity and particle physics in a quantum field theory \cite{Sugra:2021}.

Due to the coupling of supersymmetry with the translation operator, gauging translations to arrive at diffeomorphism symmetry \cite{Hehl:1976kj} implies gauging of supersymmetry \cite{Sugra:2021}, making it a local gauge symmetry. It is then usually argued that supersymmetry is broken such that diffeomorphisms as a unbroken subgroup remains. However, basically the arguments of Elitzur's theorem \cite{Elitzur:1975im,Frohlich:1981yi,Maas:2019eux} carry over to any field-theoretical path-integral approach to quantum gravity \cite{Maas:2019eux,Maas:2023emb}, implying that local supersymmetry cannot be broken. This implies that supersymmetry remains unbroken, and only manifestly supersymmetry-invariant quantities can be observed, explaining thus the absence from experiment \cite{Maas:2023emb}. Provided, of course, that supersymmetry is realized at all, it could thus be at most discovered in the same sense as the gauge symmetry of the strong interactions can be discovered.

However, as a consequence this would imply that bosons and fermions, belonging to a supermultiplet, could be as little distinguished as different colors in QCD. Apparently, this is at odds with observations. This is the next step after giving up on a physical distinction of spin states, which is already entailed by non-supersymmetric gravity \cite{Hehl:1976kj}. But this leaves the question of why these concepts appear to be so physical in particle physics terms. Conversely, this implies that any physical objects will not have any features identifying their spin, but at most their superspin.

A solution to this conundrum could be the Fr\"ohlich-Morchio-Strocchi mechanism \cite{Frohlich:1981yi,Maas:2023emb}, assuming that the Brout-Englert-Higgs effect hinted at in section \ref{s:global} would also be operative in supergravity. In gravity \cite{Maas:2019eux} it proceeds as following. Start with an invariant quantity. E.\ g., assume that there is a gravity multiplet $g$, and a matter multiplet $m$, such that an invariant quantity can be written down as $gQm$, where $Q$ is a suitable matrix (operator) to make the object invariant.

The Brout-Englert-Higgs mechanism implies that $g=\eta+h$, where $\eta$ is the dominating (classical) solution and $h$ are fluctuations. Basically, $\eta$ is the vacuum expectation value, which is a maximally symmetric solution, and, e.\ g., CDT motivates this to be the de Sitter metric. In a suitable gauge, the fluctuations $h$ will be small, such that when building a correlation function
\be
\la (gQm)(Y)(gQm)(X)\ra=\la \eta(Qm)(X)\eta(Qm)(Y)\ra+\text{ small}\label{fmsexp}
\ee
\no the first term dominates. This mechanism and expansion are well understood for electroweak physics \cite{Maas:2017wzi}, as is understood that in certain situations the small part could still have observable consequences \cite{Maas:2023emb}. In the current setting, this implies that the correlation function is dominated by the matter part, and the behavior is basically that of a matter particle in a fixed curved background.

In the same sense vertices can be constructed. Measuring spin, or a spin component, is always connected to an interaction. Already in quantum mechanics spin is measured, e.\ g., in the Stern-Gerlach experiment, by measuring the interaction with a magnetic field. Likewise, components of a superspin multiplet could be measured by a relative orientation. This is basically again like in QCD. It is impossible to measure the color of particles. But it can be measured whether a quark and an antiquark have corresponding colors or not, or whether thee quarks create a singlet or not. This is achieved by seeing an invariant object, namely if a hadron is there, or just vacuum. In the same way (super)spin components should be observable, and in the Fr\"ohlich-Morchio-Strocchi mechanism this should map on the conventional ideas of (super)spin in a fixed background.

This would explain how supergravity being a gauge theory and non-observable, could be consistent with the observation of spin. It may also pave a way to understand how experiments only discover particles and never superpartners, by being dominated by the partner and not the superpartner in expressions like \pref{fmsexp}. However, this requires to build suitable operators for this purpose, and apply the FMS mechanism. This is already in the non-supersymmetric case a formidable challenge \cite{Maas:2022lxv}, and will thus require extensive work in the future. But in the end a realization would allow to have the cake and eat it: All the good features of unbroken supergravity survive without any contradiction to experiments not observing supersymmetry.

As supersymmetry anyhow needs to be supergravity in a world with gravity, and quantum gravity seems to exhibit the necessary Brout-Englert-Higgs mechanism, this would be a very intriguing possibility indeed.

\section{Conclusions and open questions}\label{s:summary}

Only manifestly invariant quantities can describe physical observables, in quantum gravity this requires manifest diffeomorphism invariance. Using CDT, we implemented this tenet by studying a correlation function of an invariant scalar curvature measure. The results suggest that, for the sets of parameters studied, there exsists a distance regime, in which the correlator shows properties similar to what is expected for a flat-space particle. While many systematic question remain to be investigated, this is strong motivation to do so. If it can be established this implies that an object of self-bound gravitons would behave in some regime like a particle. As such objects are the only one at our disposal to describe at the quantum level objects like black holes, of which the particle-like properties have been established for some distance regime, this is an important step. Of course, and we also see this, this cannot be true at all distance scales in a de Sitter-like universe \cite{Parker:2009uva}. This immediately raises the question if a similar mechanism also applies to matter, for which quenched EDT results give already hope \cite{Dai:2021fqb}.

At the same time, we confirm that CDT in this parameter regime can be understood to exhibit a Brout-Englert-Higgs-like effect. This could allow to understand the properties of such composite objects in simpler terms, using the Fr\"ohlich-Morchio-Strocchi mechanism \cite{Maas:2019eux,Maas:2022lxv}. This would also explain a substantial amount of the simplicity observed in general relativity, which is also very well described in suitable gauges as very small perturbations around simple, exact solutions. Making this effective description precise would probably be one of the most important steps towards practical investigations using semi-perturbative methods of non-perturbative quantum gravity. Results in asymptotic saftey already support the idea that this is possible \cite{Eichhorn:2018ydy}.

Finally, taking these insights seriously, the requirement of being manifest invariant has far-reaching consequences. One example was that supergravity, which is necessarily the way how supersymmetry must manifest in a universe with gravity, implies that observables need to be manifestly invariant non-supersymmetric. In fact, the possibility of observing in such a world bosons and fermions separately is most likely only possible in a combination of a Brout-Englert-Higgs effect and the Fr\"ohlich-Morchio-Strocchi mechanism. At the same time, it explains the absence of supersymmetry in collider experiments, a tantalizing option.

\bibliographystyle{JHEP}
\bibliography{bib}

\end{document}